\documentclass[a4paper, 10pt, conference]{./ieeeconf/ieeeconf}      

\IEEEoverridecommandlockouts                              

\usepackage{graphics} 
\usepackage{epsfig} 
\usepackage{amsmath} 
\usepackage{amssymb}  
\usepackage{cite}

\usepackage{algpseudocode}
\usepackage{bbm}
\usepackage{mathtools}

\usepackage[linesnumbered,ruled]{algorithm2e}

\usepackage[caption=false,font=footnotesize]{subfig}

\begin{document}

\title{  \bf
A Machine-Learned Ranking Algorithm for Dynamic and Personalised Car Pooling Services
}

\author{Mattia Giovanni Campana, Franca Delmastro and Raffaele Bruno\\
IIT-CNR \\
Via G. Moruzzi 1, 56124, Pisa, ITALY \\
{\tt\small \{m.campana,f.delmastro,r.bruno\}@iit.cnr.it}
\thanks{*This work has been partially supported by the EC under the H2020-SC Lighthouse Project n. 691735, REPLICATE.}
}

\maketitle
\thispagestyle{empty}
\pagestyle{empty}

\begin{abstract}
Car pooling is expected to significantly help in reducing traffic congestion and pollution in cities by enabling drivers to share their cars with travellers with similar itineraries and time schedules. A number of car pooling matching services have been designed in order to efficiently find successful ride matches in a given pool of drivers and potential passengers. However, it is now recognised that many non-monetary aspects and social considerations, besides simple mobility needs, may influence the individual willingness of sharing a ride, which are difficult to predict. To address this problem, in this study we propose \textsc{GoTogether}, a recommender system for car pooling services that leverages on learning-to-rank techniques to automatically derive the personalised ranking model of each user from the history of her choices (i.e., the type of accepted or rejected shared rides). Then, \textsc{GoTogether} builds the list of recommended rides in order to maximise the success rate of the offered matches. To test the performance of our scheme we use real data from Twitter and Foursquare sources in order to generate a dataset of plausible mobility patterns and ride requests in a metropolitan area. The results show that the proposed solution quickly obtain an accurate prediction of the personalised user's choice model both in static and dynamic conditions.  
\end{abstract}

%
%
\section{Introduction}
\noindent
Car pooling (aka ride-sharing) consists in the sharing of private cars and related journeys with one or more people who have similar mobility needs. Car pooling is commonly considered a sustainable transportation mode since it reduces the number of travelling cars, which is beneficial to lower traffic congestion on roads, the need of parking spaces and total carbon emissions~\cite{TEAL1987,handke2013}. Car pooling is not a novel concept. In the past, local authorities already tried to promote ride-sharing for commuters, starting with the construction of high-occupancy vehicle lanes in early 1980s. However, only recently car pooling started to gain momentum through the development of online and mobile services that allow drivers with spare seats to connect with people wishing to share a ride on very short notice or even en-route (e.g., BlaBlaCar, carpooling.com, gomore.com).   

In order to be successful, car pooling applications need efficient \emph{matching} algorithms able to automatically provide suitable and real-time ride matches to their users~\cite{Agatz2012}. Typically, proximity in time and space is a necessary condition to have a good match between trips~\cite{Trasarti2011_kdd,TITS14_he}. Clearly, private car-pooling providers want to generate revenues and maximise the number of participants. Public providers may also have a societal objective and aim at maximising a system-wide benefit (e.g., reduction of congestion). Thus, when determining matches between drivers and riders in a
ride-sharing system, it is essential to effectively combine system-wide optimisations with user-based benefits and constraints on the feasibility of ride matches.

It is important to recognise that reduced travel costs may not necessarily be the only or most important reason for a user to accept a ride-sharing suggestion~\cite{TEAL1987}, especially in case of short distances. Many other aspects may be relevant for the user's choice, and determine whether a particular shared ride would be accepted or not (e.g., safety considerations, social similarity between driver and passengers, etc.). For these reasons, many recommendation systems and incentive models have been recently proposed to increase the success probability of ride-sharing suggestions, for instance on the basis of monetary negotiation~\cite{taai11_cho}, measurements of ride enjoyability~\cite{itsc15_guidotti} and utility of the user's desired activity at the destination~\cite{itsc15_lira}. However, the majority of existing solutions assume to know \emph{a priori} the most relevant reasons to accept or reject a shared trip for each user, typically on the basis of stated-preference travel surveys~\cite{Correia2011}. Furthermore, users' preferences may change over time making the users' profiles difficult to maintain.  
 
In this work we propose \textsc{GoTogether}, a \emph{dynamic and personalised car pooling solution that is able to learn the individual acceptance model} of each user in an automated and transparent (for the user) way. We start by observing that any online car pooling system provides the passengers with an \emph{ordered} list of the top ride matches to choose from. The user can accept one of the suggested offers (not necessarily the top ranked) or reject all of them. The user's choices over time provide invaluable information on her personal preferences. For this reason, we leverage on \emph{machine-learned ranking} (also Learning-to-Rank or LR) techniques~\cite{Liu2009} to reconstruct the initially unknown ranking model that is implicitly adopted by each individual user to determine the \emph{relevance} of a ride match for a specific request of the user. Then, \textsc{GoTogheter} builds a personalised list of recommended shared rides for each user in order to maximise the success rate of the offered ride matches. 
To investigate the effectiveness of the proposed solution we used a \emph{data-driven} validation methodology generating a data set that merge topological information with the social characteristics of the visited places and of possible car poolers. To this aim, we extracted data from FourSquare and Twitter online social networks as explained in Section ~\ref{sec:results}. The results show that the proposed solution can obtain an accurate prediction of the personalised user's choice model after a few replications of the same car pooling requests. Furthermore, our learning algorithm quickly reacts to variations of the users' profiles and dynamically adjust the users' ranking models. 

The rest of this paper is structured as follows. Section~\ref{sec:related} provides an overview of related work. Section~\ref{sec:pooling} presents \textsc{GoTogether} and the proposed learning framework. In Section~\ref{sec:results}, we present numerical results for the analysed case study. In Section~\ref{sec:prototype}, we describe GoTogether mobile application, currently in use for a pilot testing. Finally, in Section~\ref{sec:conclusions} we draw our conclusions and present directions for future research. 
%
%
%
%
%
\section{Related Works}\label{sec:related}
\noindent
There is large body of work on the carpooling problem. A thread of studies focuses on determining the potential of carpooling services in urban transportation scenarios mining big mobility data. For instance, authors in~\cite{Trasarti2011_kdd} estimate the percentage of sharable traffic for private cars in Tuscany by extracting mobility profiles and route similarity between routine trips from GPS-based car trajectories. In~\cite{Santi2014}, the benefits of vehicle pooling for taxi services in New York is quantified as a function of tolerable passenger discomfort. Mobile and online social data is used in~\cite{Cici2014} to assess the potential of ride-sharing for reducing traffic in the cities of Barcelona and Madrid. All the aforementioned studies show that a range from 30\% to 70\% of existing trips can be typically shared. 

Many carpooling works are related to the design of efficient algorithms for matching passengers and drivers with similar mobility needs, and scheduling riders' pickup and delivery, in order to maximise the benefits of carpooling (e.g., minimising the total travelled distances or maximising the number of carpoolers) considering a range of constraints and rider preferences (e.g., maximum waiting time or social distance). A survey of optimisation frameworks for the dynamic carpooling problem can be found in~\cite{Agatz2012}. For instance, integer programming is used in~\cite{Baldacci2004} to solve the carpooling problem. Genetic algorithms are proposed in~\cite{Herbawi2012,Huang15} to reduce computational times. Frequency-correlated algorithms for rider selection and route merging are developed in~\cite{TITS14_he}. A stochastic carpooling model that considers the influence of stochastic travel times is formulated in~\cite{TITS14_yan}.  

Recently, other studies focus on designing recommendation systems to improve the acceptance probability of a carpooling match and to encourage participants to use the carpooling service. The authors in~\cite{Yan2011} develops a model for the carpooling problem that incorporates pre-matching information (e.g., previous accepted passengers). Network analytics is used in~\cite{Guidotti2016} to determine subpopulations of travellers in a given territory with a higher change to create a carpooling community, and the predisposition of users to be either drivers or passengers in a shared car. A measure of enjoyability for a carpooling ride is defined in~\cite{itsc15_guidotti} based on social similarities between any two users and tendency of a person to group with similar ones. In~\cite{Chang2011} an route planning algorithm is proposed to generate the top-$k$ personalised routes based on route familiarity for each user. Our work differs from the aforementioned studies because we leverage on the history of user's interactions with the carpooling system to incrementally \emph{learn} the acceptance model of each user. 

%
%
%
%
%
%
\section{\textsc{GoTogether}: a Dynamic and Personalised Car Pooling Service\label{sec:pooling}}
\noindent
In this section we describe the system architecture of \textsc{GoTogether} and we present its core functionalities, focusing on the learning algorithm used to infer the users' personal ranking model.
%
%
%
%
\subsection{System architecture}
\begin{figure}[t]
    \centering
    \includegraphics[width=0.45\textwidth]{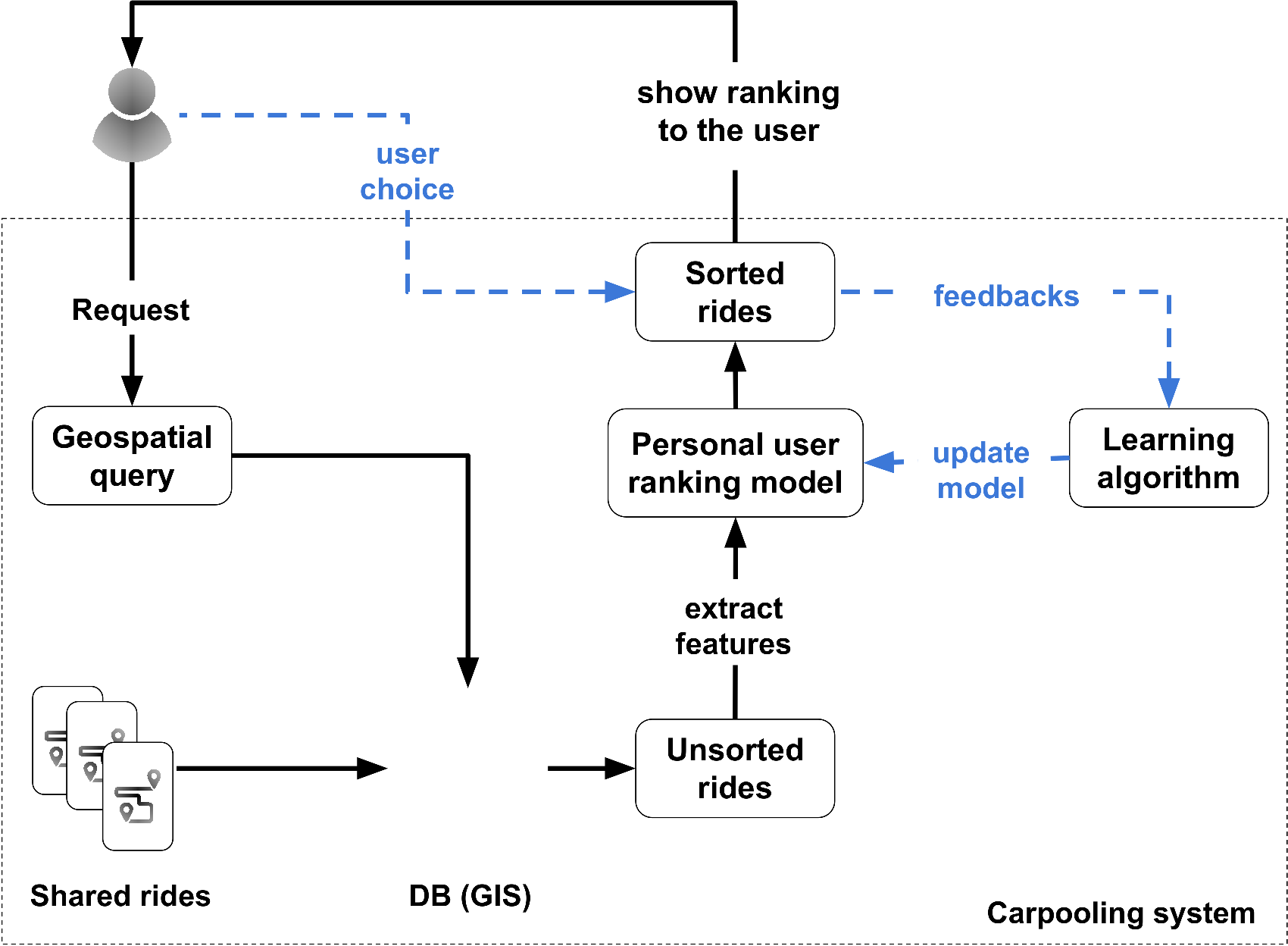}
    \vspace{-0.4cm}
    \caption{GoTogether system architecture.}
    \label{fig:if_ml}
    \vspace{-0.7cm}
\end{figure}

\noindent
Figure~\ref{fig:if_ml} illustrates the system architecture of \textsc{GoTogether}, highlighting the operation flows between the user and the system during the ride selection process. The basic component of the system is a spatial database that stores all the offered trips. A passenger's query for a shared trip triggers the ride searching process, which generates a list of possible ride matches. Then, the candidate trips are ranked according to the estimated user' ranking model in order to maximise the success probability of a ride match. The passenger's query must provide a series of parameters to define the ride search. Specifically, it is necessary to specify at least: i) the departure place ($q_{sp}$), ii) the destination place ($q_{dp}$), and iii) the desired departure time ($q_{dt}$). Typically, the query can also include the user's preferences for the ride, such us the tolerance for pick-up/drop-off distances, the tolerance for the deviation from the preferred departure time, and desired trip and driver's characteristics.

\begin{figure}[t]
    \centering
    \includegraphics[width=0.4\textwidth]{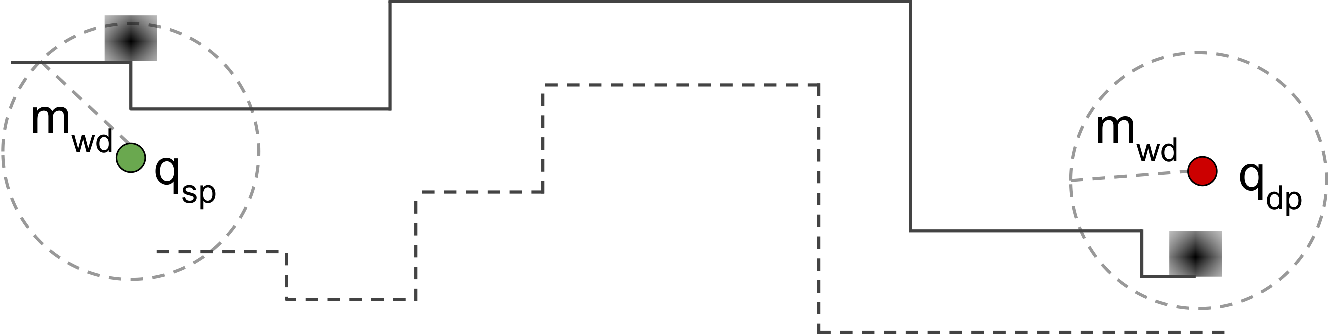}
    \vspace{-0.4cm}
    \caption{Example of selection of a candidate ride.}
    \label{fig:rides}
\vspace{-0.6cm}
\end{figure}

To obtain the list of candidate shared trips \textsc{GoTogether} applies the following procedure. First of all, it defines the \emph{pickup area} and \emph{drop-off area} of the potential passenger as the circles of radius $\delta$ around the $q_{sp}$ and $q_{dp}$ points, respectively\footnote{$\delta$ is a system parameter that defines the maximum walking distance from the passenger's departure/arrival locations to the pickup/drop-off points.}. In addition, we denote with $\tau$ the maximum delay of the shared trip with respect to the desired departure time. Then, for each retrieved ride in the database, say $r_{i}$, we compute the shortest paths between $q_{sp}$ and $q_{dp}$. The intersections between the shortest paths originated from $q_{sp}$ and $q_{dp}$ and $r_{i}$ are the pickup points and drop-off points of the passenger, respectively. The pick-up delay is obtained as the difference between the desired departure time of the passenger and the time instant at which the driver reaches the pickup point following $r_{i}$. Finally, $r_{i}$ is a candidate ride match for the passenger's query if the pickup and drop-off points fall within the pickup and drop-off areas, and the pick-up delay is shorter than $\tau$. Figure~\ref{fig:rides} illustrates an example of the above-described ride selection process for a given request (solid line is a candidate ride, while dashed line no).  

The list of candidate rides extracted from the ride database needs to be ordered based on the passengers' preferences. To this end, the user's \emph{personal ranking model} (also called \emph{ranker}) is applied to this list to assign a ranking score to each shared trip. Typically, this score is obtained by a combination of \emph{utility functions} associated with a set of ride features. Clearly, the system does not have the complete and exact knowledge of the user's ranker but it has to rely on an estimated model. In this study, \emph{we advocate the use of the history of users' choices to predict the users' rankers}. Specifically, we leverage on LR techniques for automatically learning the ranking model, and therefore optimise the car pooling recommendations. As better explained in Section~\ref{sec:ranking}, when the user accepts a ride from the proposed ranking, the system generates a training data which is then used by the learning algorithm to produce the ranking model. 

%
%
%
%
\subsection{The ranking model \label{sec:ranking}}
\noindent
Before describing the \textsc{GoTogether} learning algorithm of the individual ranking models, it can be useful to provide a brief overview of Learning-to-Rank (LR) techniques. 
%
%
%
\subsubsection{Background on Learning-to-Rank\label{sec:LR}}
\noindent
Learning-to-Rank (LR) was originally proposed for Information Retrieval (IR) systems, i.e., collections of data objects (text documents, images, trajectories, etc.), which can be queried by multiple users to obtain ranked lists of objects that match the queries with different degrees of relevance. Then, machine learning techniques can be applied to IR systems in order to automatically discover the users' ranking models~\cite{hang2011short}. Most of LR methods	employ \emph{offline} supervised learning approach, i.e., rankers are estimated before deploying the IR system using training data that has been created in advance~\cite{liu2009learning}. This approach has two main drawbacks: (i) it requests a large amount of manually annotated data (i.e., the \emph{training} and \emph{test} sets) needs to be available before deployment, and (ii) it is difficult and costly to track dynamic behaviours in a timely manner. On the contrary, \emph{online} LR techniques allow the system to learn directly from the users' interactions, e.g. via click actions\footnote{Typically, IR systems are web-based and a click corresponds to the user's choice of a data object in the ranked list or to an expression of interest in a specific data object.} ~\cite{liu2009learning}. This type of solutions are typically based on reinforcement learning techniques, meaning that the system test new rankers, and learns from users' interactions with the presented rankings. We believe that the online approach is best suited for a car pooling system, since collecting a large amount of training data before the system's deployment is not feasible. Furthermore, car pooling users may show dynamic behaviours, and the rankers should be able to self-adapt during the system lifetime. 

Two of the most successful approaches to LR are the \emph{listwise} and \emph{pairwise} methods, which differentiate on the basis of the type of users' feedbacks and cost functions used to evaluate the performance of the learned ranking functions. More precisely, listwise approaches directly operate on the entire ranked list of data objects associated with a query. In pairwise approaches the learning procedure consider as input pairs of objects, and it assigns a label to the pair representing the relative relevance of the two objects for the user. In this case the LR method learns a classifier that predicts these labels for each possible pair of data objects in the query result. We believe that pairwise LR techniques fits better a car pooling system because each query generates a single output, i.e., the selected trip. Thus, for a pairwise LR approach, it is easier to generate a sufficiently large sequence of training data from a single query, while the system is running.

Finally, it is important to point out that online LR methods intrinsically suffer from the \emph{exploitation-exploration dilemma}. In other words, an LR algorithm needs to both explore new solutions to obtain feedback for effective learning, and exploit what has already been learned to produce results that are acceptable for the users. A well-known method for balancing exploration and exploitation is the $\epsilon$-greedy strategy~\cite{watkins1989learning}, in which the agent selects at each time step the \emph{greedy action}\footnote{In \textsc{GoTogether} an action is the selection of a ride match.} (i.e., the action with the highest currently estimated value) with a constant probability $1-\epsilon$, and a random action with probability  $\epsilon$.  However, in~\cite{silverstein1999analysis} it has been shown that implicit feedback can be biased towards the top results displayed to the user. The user may not choose the most relevant ride simply because it is located in the lower section of the proposed list.
%
%
%

\subsubsection{The learning algorithm\label{sec:LR_algorithm}}
\noindent
\textsc{GoTogether} uses an online and pairwise LR approach to define the learning algorithm, which is inspired by the technique developed in~\cite{hofmann2013balancing}. Our algorithm, which is described in Algorithm~\ref{algo:pairwisesgd}, takes as input a user $u$, the set of candidate rides $R$ fetched from the database with a randomised order for a specific query, the learning rate $\eta$, and the probability $\epsilon \in [0,1]$. As better explained in the following, $R$ is the \emph{explorative list} of ride matches because rides are not yet sorted based on their relevance and their position in the list is random.
The algorithm starts by extracting the vector of features $\mathbf{x} = \boldsymbol{\phi}(r)$ from each candidate ride $r \in R$. The set of features used in this study to rank the potential ride matches is explained in Section~\ref{sec:ranking}, but it can be further extended. We associate a weight $w_{k}$ with each feature $x_{k} \in \boldsymbol{\phi}(r) $. Then, the candidate rides are ranked using a weighted linear combination of these features. Specifically, the estimated user's ranker at time step $t$ corresponds to the vector of ranking weights, say $\mathbf{w_{t-1}}$, learned so far. Then the learning algorithm construct the \emph{exploitative list} $L$ by sorting the list $R$ of candidate rides using the estimated ranker. Finally, a \emph{recommendation list} $I$ is selected from $L$ and $R$ as follows. For each ranking position, the algorithm selects the corresponding ride from the exploitative list $L$ with probability $1-\epsilon$; otherwise, with probability $\epsilon$, the algorithm selects a ride from the explorative list $R$.

\setlength{\textfloatsep}{0pt}
\begin{algorithm}[t]
	\small
    \KwIn{$u$, $R$, $\eta$, $\epsilon$}
    
    $R = fetchCandidateRides(q_t, u)$
    
    $\mathbf{X} = \phi(R)$ // extract features
    
    $\mathbf{w_{t-1}} = fetchUserRanker(u)$
    
    // construct exploitative result list
    
    $S = \mathbf{w_{t-1}^T X}$
    
    $L = sortDescendingByScore(R, S)$
    
    $I[r] \leftarrow $ first element of $L \notin I$ with probability $\epsilon$; element randomly sampled without replacement from $L \setminus I$ with probability $1-\epsilon$
    
    Display $I$ to u and observe accepted ride $r_s$.
    
    Construct labeled pairs $P = (\mathbf{x_{a}}, \mathbf{x_{b}}, y)$ from $I$ and $r_s$.
    
    // update model

        \For{$i$ in $1$\ldots$P$}{

            \If{$y_i (\mathbf{x_{a\_i}} - \mathbf{x_{b\_i}}) \mathbf{w_{t-1}^T} < 1.0$ and $y_i \neq 0.0$}
            {
            
            $\mathbf{w_t} = \mathbf{w_{t-1}} + \eta y_i (\mathbf{x_{a\_i}} - \mathbf{x_{b\_i}})$
            }
        }

    \caption{The carpooling learning algorithm.}
    \label{algo:pairwisesgd}
\end{algorithm}

\vspace{-0.1cm}
At this point, the system shows the resulting recommendation list to the user, and it observes the user's feedback. Two types of feedbacks are possible. On the one hand, the user can reject the entire recommendation list if the relevance of all shown results is too low (i.e., below a critical relevance threshold). On the other hand, the user can accept one of the proposed rides, not necessarily the one ranked first. If the user accepts a ride, the algorithm infers all the possible labeled ride pairs $P$ using the pairwise labelling method described hereafter. For the sake of presentation clarity, we introduce the operator $\succ$, and $r_i \succ r_j$ means that the ride $r_i$ is more relevant than the ride $r_j$ for the user. Let us assume that the recommendation list that is showed to the user contains four ride matches $(r_1, r_2, r_3,r_4)$ and the user accepts ride $r_3$. Then, we can infer that $r_3 \succ r_1$, $r_3 \succ r_2$, and $r_3 \succ r_4$, but we can not say anything about the relevance between $r_1$ and $r_2$. From these observations, three training pairs can be obtained as $(r_1, r_3, -1)$, $(r_2, r_3, -1)$, and $(r_3, r_4, +1)$, where the labels ``$-1$'' and `$+1$'' mean that they are negative and positive learning instances, respectively. In other words, the learning algorithm should update the user's ranker in order to prefer a ride similar to $r_3$ (assigning a higher rank to it) than a ride like $r_2$ or $r_3$ the next time the user makes a similar query $q$. More formally, for each training pair $(\mathbf{x_{a}}, \mathbf{x_{b}}, y)$ in the training data $P$, the algorithm measures how much the current model has mis-labeled the examples. If the labels don't match, the weight vector is updated with the unregularized \emph{Stochastic Gradient Descent}~\cite{zhang2004solving} update rule:
\begin{equation}
	\small
	\nonumber
	\mathbf{w_t} = \mathbf{w_{t-1}} + \eta \;  y_i (\mathbf{x_{a}} - \mathbf{x_{b}}),
	\label{eq:sgd_update}
\end{equation}
where $\mathbf{x_a}$ and $\mathbf{x_b}$ are the features vectors of the ride pair. The update rule adjusts the model weights in order to minimise the number of mis-labeled pairs. The parameters $\eta$ influences the rate of learning but also the convergence speed of the learner and its tuning is essential to avoid excessive fluctuations of the learner weights.

\begin{figure}[t]
    \centering
    \includegraphics[width=0.43\textwidth, trim={3.9cm 0cm 6.7cm 0cm},clip,angle=0]{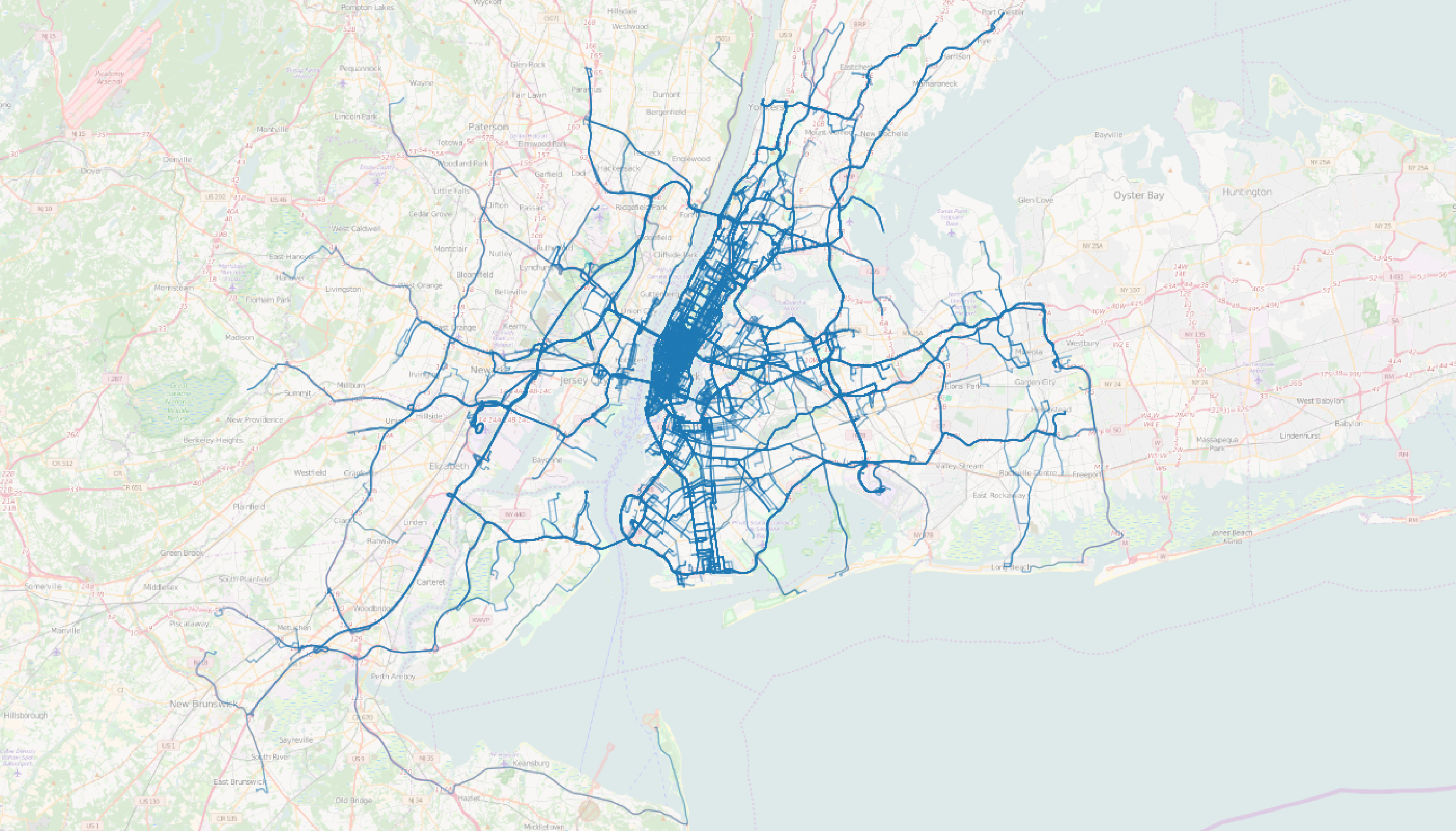}
    \caption{The trip trace in New York, inferred from the Foursquare's check-ins shared in Twitter.}
    \label{fig:ny_traces}
\end{figure}

\section{Experimental Evaluation}\label{sec:results}
\noindent
In order to assess the performance of our learning algorithm we generate synthetic users and mobility traces using real-world data sources. Our dataset, evaluation methodology and experimental results are described in the following sections.
%
%
%
%
%



%
%
%
\subsection{Data sources\label{sec:data}}
\begin{figure}[tbhp]
	\vspace{-0.3cm}
    \centering
    \includegraphics[width=0.4\textwidth]{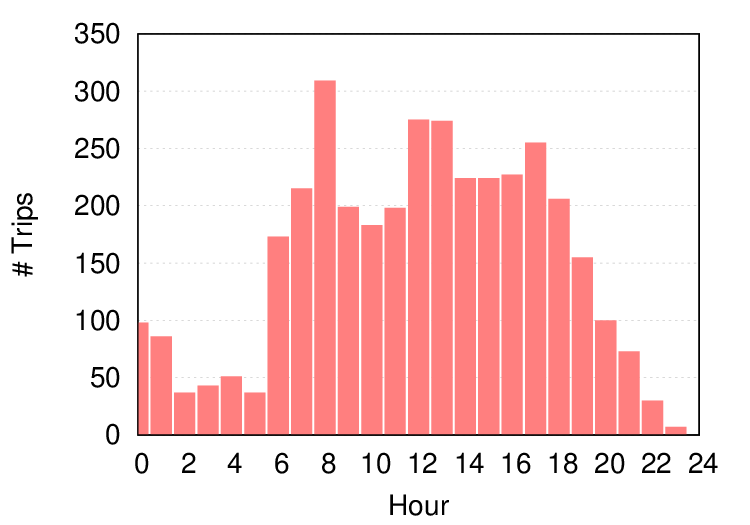}
    \vspace{-0.35cm}
    \caption{Hourly distribution of trips in our mobility trace.}
    \label{fig:rides_hour}
\vspace{-0.6cm}
\end{figure}

\noindent
Nowadays, Online Social Networks (OSNs) can be effectively used to study different aspects of human behaviours, as well as to obtain information regarding individual mobility patterns. In this study we jointly use Twitter and Foursquare as data sources~\cite{wang14_osn}. Specifically, Foursquare is a location-based OSN that motivates registered users to share their check-ins at different places. A check-in is often characterised not only through raw GPS coordinates, but also with contextual information such as the location name (e.g., ``Starbucks'') and its semantic description (e.g., coffee shop). Foursquare does not provide an API to fetch the check-ins generated in a given geographic area. However, Foursquare users typically share their check-ins also with other OSNs like Twitter. Furthermore, Twitter provides supplemental information about social connections and interest similarities between users. 
The following methodology is used to obtain the dataset for our experiments. First, we leverage on the Twitter streaming APIs to get a set of geolocated tweets sharing Foursquare check-ins within the metropolitan area of New York for two weeks at the end of February 2016. In this way, we collect the check-ins of 56 users. For each user, we also download the tweet history and we use the TagMe annotation tool~\cite{ferragina2010fast} to extract the users' topics of interest\footnote{We remind that Twitter APIs allow to freely download only the last 3200 tweets of a user.}. Finally, we employ Foursquare APIs to expand the set of topics of each user with the semantic categories of his check-ins. To infer a plausible mobility traces from the users' check-ins, we proceed as follows. First, for each user in our trace we aggregate all check-ins in a single day and we sort them by their timestamp. Then, we use the \emph{Google Maps Directions APIs} in order to determine the most plausible car trajectory between pairs of consecutive locations in each day. Finally, we prune all the trips with a duration shorter than 20 minutes, which results in maintaining a total of 3679 trips. Figure~\ref{fig:ny_traces} shows the spatial distribution of these trips on the analysed geographical area, while Figure~\ref{fig:rides_hour} shows the hourly distribution of the trips over a day. Typical peak and off-peak behaviours can be observed. 
%
%
%
\subsection{User's choice model\label{sec:choice_model}}
\noindent
In the transportation field various discrete choice models have been proposed to characterise the probability of individuals choosing a given transportation option from a finite set of alternatives~\cite{book_transport11}. To represent the attractiveness of the alternatives, the concept of utility is typically used, and the observable utility is usually defined as a linear combination of features associated to each transportation alternative. Furthermore, a weight is associated to each feature to quantify the relevance of that feature for an individual.

In this study, we use the following four features to rank a ride offer:
\begin{list}{\tiny$\bullet$}{\leftmargin=1em \itemindent=-0.5em}
\item the \emph{walking distance} from the trip origin to the pickup point ($d_p$);
\item the \emph{walking distance} from the drop-off point to the trip destination ($d_d$);
\item the \emph{pickup delay} ($t_p$);
\item the \emph{social similarity} between the driver and the passenger.
\end{list}
It is intuitive to recognise that walking distances may have different degrees of utility for each user. In general, the shorter the walking distance and the higher the utility. To represent this variability, we describe the walking distance as the combination of three features, which correspond to three non-overlapping distance ranges. Specifically, ranges $[0,d^1]$, $({d^1},d^2]$ and $({d^2},d^3]$ correspond to short, medium and long walking distances, respectively. Then, a weight $\omega^{1} (d_p)$, $\omega^{2} (d_p)$, $\omega^{3} (d_p)$ for the walking distance from the trip origin to the pickup point, and $\omega^{1} (d_d)$, $\omega^{2} (d_d)$, $\omega^{3} (d_d)$ for the walking distance from the drop-off point to the trip destination, are assigned to each one of the previous ranges, respectively. Similarly, we model the pickup delay as the combination of three features, which correspond to three non-overlapping time ranges. Specifically, ranges $[0,t^1]$, $({t^1},t^2]$ and $({t^2},t^3]$ correspond to short, medium and long delays, respectively. Then, a weight $\omega^{1} (t_p)$, $\omega^{2} (t_p)$, and $\omega^{3} (t_p)$ is assigned to each one of the previous ranges, respectively\footnote{In the following experiments $d^1=1$~Km, $d^2=2$~Km and $d^3=3$~Km. Similarly, $t^1=30$~minutes, $t^2=60$~minutes and $t^3=90$~minutes.}. 

The fourth feature is a measure of the common interests between users, as in~\cite{itsc15_guidotti}. Specifically, for each pair of users $u$ and $v$ we can build two vectors of topics, say $\vec{t}_u$ and $\vec{t}_v$, from their tweets, where each topic is weighted by its relative importance (i.e., frequency) within the tweets. The similarity between these two vectors is estimated using the cosine similarity, i.e., the cosine of the angle between the vectors of topics:

\begin{equation*}
\vspace{-0.1cm}
\small
sim(\vec{t}_{u},\vec{t}_{v}) = \frac{\vec{t}_{u}\cdot \vec{t}_{v}}{||\vec{t}_{u}|| || \vec{t}_{v} ||} \; 
\end{equation*}
From the social similarity we can also derive the \emph{homophily} of user $u$, say $h_{u}$, which is defined as the median of the social similarity between this user and all his friends. If $h_u \approx 1$, we say that $u$ is \emph{homophilous}, while if $h_u \approx -1$ we call $u$ \emph{heterophilous}. In the former case, the user tends to associate and bind with similar others, while in the latter case with individuals that have different interests. Thus, we expect that this property may also influence users' choices of attractive ride shares. Clearly, varying degrees of homophilous and heterophilous behaviours can be identified. 

Finally, we can express the total utility, for a user $u$, of a ride $r$ offered by driver $v$ as follows:
\begin{equation}
	\vspace{-0.2cm}
	\small
	\label{eq:utility}
	\begin{split}
		\tiny
	    U_u(r) {}&= h_u \cdot sim(\vec{t}_{u},\vec{t}_{v}) + \sum_{j = 1}^3 \omega^{j} (t_p) \cdot \mathbbm{1}\{ t_p \in [t^{(j-1)},t^j] \}	\\
	  	& + \sum_{x= p,d} \sum_{j = 1}^3 \omega^{j} (d_x) \cdot \mathbbm{1}\{ d_x \in [d^{(j-1)},d^j ]\},
	\end{split}
\end{equation}
where $\mathbbm{1}\{z\}$ is the indicator function of a condition $z$: $\mathbbm{1}\{z\} = 1$  if $z = true$, and $\mathbbm{1}\{z\} = 0$ otherwise. In other words, for the sake of simplicity the utility associated with each feature is equal to one, but different weights are assigned to each feature. It is important to note that we do not need to learn the utility functions but only their weights. 


Based on the values of the weights we have defined four categories of typical users:
\begin{list}{\tiny$\bullet$}{\leftmargin=1em \itemindent=-0.5em}
\item \textbf{Homophilous and lazy users ($U_1$)}. They have a high level of homophily and they prefer rides with a short walking distance, and a short pickup delay: $h_i = 0.9$; $\omega^{1} (t_p) = \omega^{1} (d_p) = \omega^{1} (d_d) = 0.8$; $ \omega^{2} (t_p) = \omega^{2} (d_p) = \omega^{2} (d_d) = 0.15$;  $ \omega^{3} (t_p) = \omega^{3} (d_p) = \omega^{3} (d_d) = 0.05$.
\item \textbf{Homophilous and active users ($U_2$)}. They have a high level of homophily and they are willing to walk longer distances to reach the driver, and wait a longer time: $h_i = 0.9$; $\omega^{1} (t_p) = \omega^{1} (d_p) = \omega^{1} (d_d) = 0.05$; $ \omega^{2} (t_p) = \omega^{2} (d_p) = \omega^{2} (d_d) = 0.15$;  $ \omega^{3} (t_p) = \omega^{3} (d_p) = \omega^{3} (d_d) = 0.8$.
\item \textbf{Heterophilous and lazy users ($U_3$)}. They have a low level of homophily, and they prefer a short walking distance and a short pickup delay: $h_i = 0.1$; $\omega^{1} (t_p) = \omega^{1} (d_p) = \omega^{1} (d_d) = 0.8$; $ \omega^{2} (t_p) = \omega^{2} (d_p) = \omega^{2} (d_d) = 0.15$;  $ \omega^{3} (t_p) = \omega^{3} (d_p) = \omega^{3} (d_d) = 0.05$.
\item \textbf{Heterophilous and active users ($U_4$)}. They have a low level of homophily, and they are willing to walk a long distance and to wait a longer time: $h_i = 0.1$; $\omega^{1} (t_p) = \omega^{1} (d_p) = \omega^{1} (d_d) = 0.05$; $ \omega^{2} (t_p) = \omega^{2} (d_p) = \omega^{2} (d_d) = 0.15$; and $ \omega^{3} (t_p) = \omega^{3} (d_p) = \omega^{3} (d_d) = 0.8$.
\end{list}
%
%
%
%
\subsection{Evaluation methodology and results}
\noindent
To test the performance of the proposed car pooling system we use the following methodology. First, we assume that the users in our dataset are \emph{commuters}, who perform the same set of ride-sharing requests over several consecutive days. Then, we uniformly distribute the users in the previously described categories (i.e., $U_1$, $U_2$, $U_3$, and $U_4$). 

\begin{figure}[tbhp]
	\vspace{-0.4cm}
    \centering
    \includegraphics[width=0.4\textwidth]{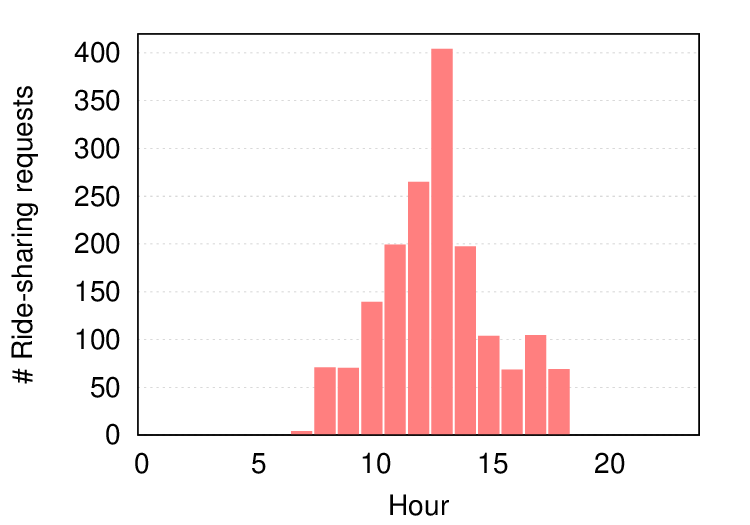}
    \vspace{-0.3cm}
    \caption{Hourly distribution of the average number of feasible queries generated by each user.}
    \label{fig:queries_hour_avg}
    \vspace{-0.3cm}
\end{figure}

The requests of shared rides for each user are generated as follows. We consider the mobility trace of each user in the dataset and we cluster both the origin and the destination points of the trips in the trace. We assume that two points belong to the same cluster if the distance between them is shorter than 400~meters. Then, we use the centroids of these clusters as the origin and destination points of the queries performed by that user. To avoid searching for unpopular and short routes, we also require that the requested ride is not shorter than 10~Km, and that there are at least fifteen ride matches for that query in the mobility database. To assess the load of our car pooling service, Figure~\ref{fig:queries_hour_avg} shows the average number of feasible requests generated by each user on a hourly basis. As expected, the car pooling service has a load peak in the middle of the day. Clearly, the number of feasible requests is varying between the users. To avoid a bias towards users that are much more active than others, we randomly select at most 100~queries per hour for each user from the set of feasible ride-sharing requests. Finally, we assume that the recommendation list consists of \emph{ten} suggested ride matches, and that the user selects one of the recommended rides only if the ride utility, as defined in Equation~(\ref{eq:utility}), is greater than a critical threshold called $C$. 

Considering all the feasible rides, the average ride utility per user varies between $0.1$ and $2.77$. However, the utility values are more concentrated in the lowest part of this range. For instance, only  $4.26\%$ of ride offers has an utility that is greater than $2$. For this reason, it does not seem reasonable to select high values of the threshold $C$. Consequently, to evaluate our system we consider three different acceptance thresholds, namely  $C =0,1,2$. Clearly, if $C=0$ then users will always select one of the proposed ride matches in the recommended list. The larger the $C$ values, the higher the probability to reject an offer. 

%
%
%
%
\subsubsection{Metrics}
\noindent
We evaluate \textsc{GoTogether} in terms of two performance metrics. The first one is the average ranking of the best ride match of each query. In principle, an ideal ranker should always classify the best ride match as the top ranking in the recommended ride list. The second one is the success probability of the ride match, computed as the ratio between the number of rejected ride requests (i.e., recommended ride lists without acceptable offers) and the total number of requests. 
\begin{figure*}[tbh] 
\centering
    \subfloat[$C=0$\label{fig:static_c0}]{%
      \includegraphics[width=0.32\textwidth]{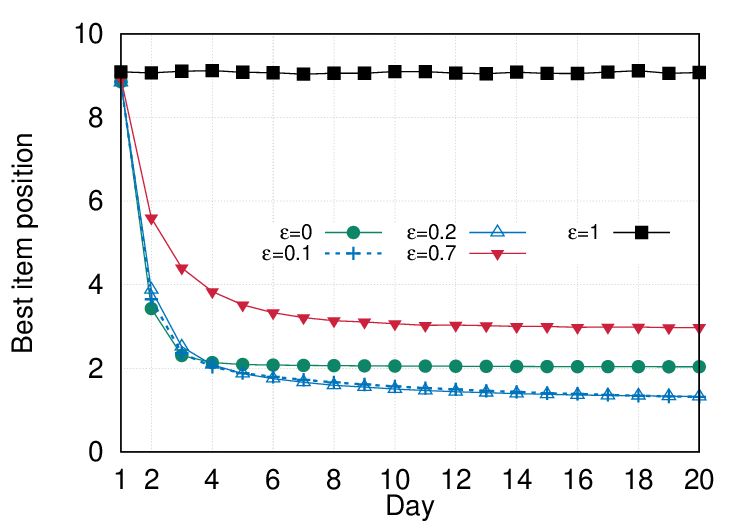}
    }
\hfill
    \subfloat[$C=1$\label{fig:static_c1}]{%
      \includegraphics[width=0.32\textwidth]{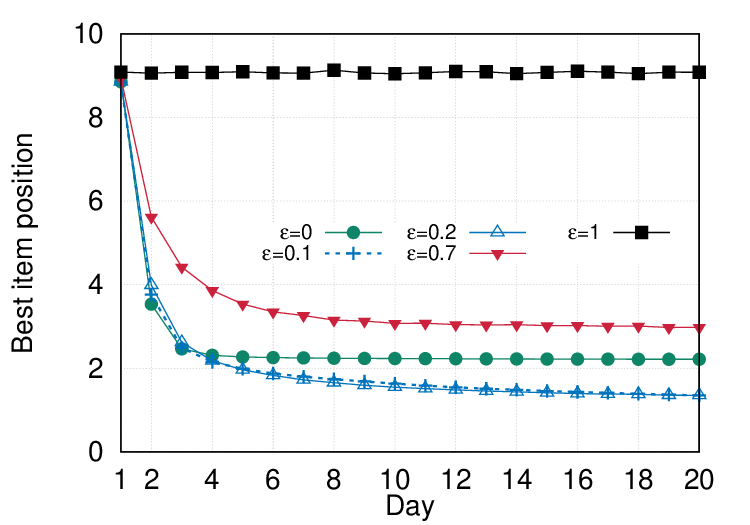}
    }
\hfill
    \subfloat[$C=2$\label{fig:static_c2}]{%
      \includegraphics[width=0.32\textwidth]{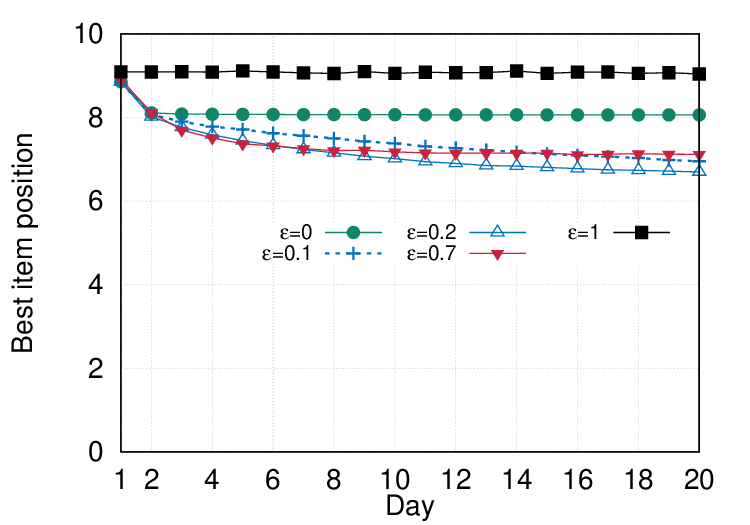}
    }
\caption{Average ranking of the best ride match for various acceptance thresholds and exploration rate.}
\label{fig:static_c}
\vspace{-0.6cm}
\end{figure*}
%
%
%
%
%
%
\begin{figure}[b]
    \centering
    \includegraphics[width=0.4\textwidth]{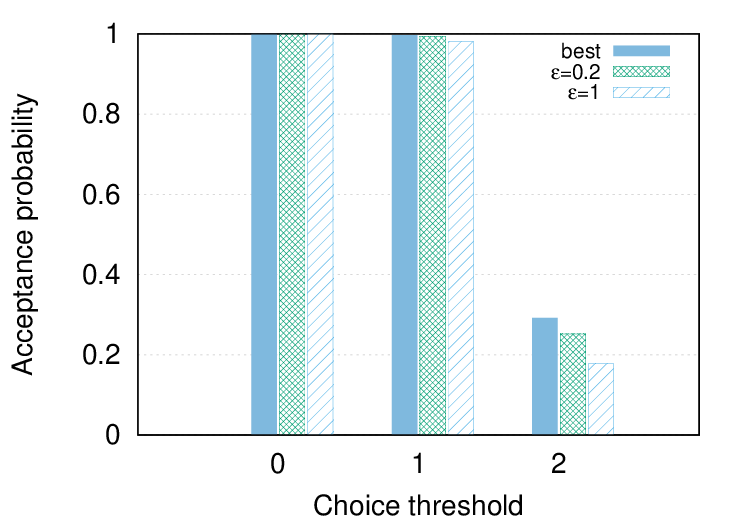}
    \vspace{-0.3cm}
    \caption{Average success probability of a ride request.}
    \label{fig:acceptance}
\end{figure}

\subsubsection{Static scenario}
\noindent
The first set of experiments is carried out in a static scenario in which each user is characterised by a choice model with time-invariant parameters. Then, we evaluate the convergence time of the learning algorithm as a function of the exploration rate $\epsilon$. Figures~\ref{fig:static_c0},~\ref{fig:static_c1}, and~\ref{fig:static_c2}  show the average ranking in the recommendation list of the best ride match for acceptance thresholds equal to 0, 1, and 2, respectively. Note that we do not assume any a priori knowledge of the users' choice model, and the users' rankers are initialised with all weights set to 0. Important observations can be derived from the shown results. First, our learning algorithm quickly improves its predicting performance and after a few iterations (i.e. days) it is able to classify the best ride as one of the top rankings in` the recommendation list. Clearly, the convergence speed to a stationary behaviour depends on the exploration rate $\epsilon$. Generally, the lower the exploration rate, the better the learning performance. Intuitively, this can be  explained by observing that in a static scenario the users apply always the same choice model and the ranker continues to learn and adapt to the users' profile. Thus, exploitative actions should be preferred over explorative actions. For instance, a purely random strategy that selects only explorative actions (i.e. $\epsilon = 1$) is unable to learn the users' ranking models and it could even fail to include the best ride match in the list of the ten recommended rides. On the other hand, $\epsilon=0.2$ and $\epsilon=0.1$ provide the highest rankings for the best ride match. Interestingly, a strategy that selects only exploitative actions (i.e. $\epsilon = 0$) performs worse than a strategy that still allows an exploration phase. Furthermore, the learning performance significantly degrades when increasing the value of the acceptance threshold $C$. In particular, with $C=2$ the learning the best ride match is classified at most with the sixth ranking even with the best setting of the exploration rate. A possible explanation of this behaviour is that there are many rejected offers and the learning algorithms has too few examples to learn from. 

To validate our intuition about the learning degradation, in Figure~\ref{fig:acceptance} we show the average success probability of a ride request for three different cases: $i)$ an ideal ranker that always classifies the best ride match at the top of the recommended list, $ii)$ the fully explorative learning algorithm, and $iii)$ the learning algorithm with the best setting of the exploration rate (i.e., $\epsilon=0.2$). We can observe that, when $C=2$, even an ideal learning algorithm would obtain a low success probability (around $28\%$). The performance of our exploitative learning algorithm with $\epsilon=0.2$ is close to that of the ideal solution, even if the best ride match is not top ranked (see Figure~\ref{fig:static_c2}). On the contrary, a random choice of the recommended list (i.e., $\epsilon=1$) leads to worse performance with a $30\%$ decrease of the success probability. In general, the learning algorithm needs the users' choices to improve its estimate of the users' ranking models. If many ride requests fail, then the learning algorithm gets stuck with inaccurate estimates.   
%
%
%
%
%
%
%
\begin{figure}[b]
    \centering
    \includegraphics[width=0.4\textwidth]{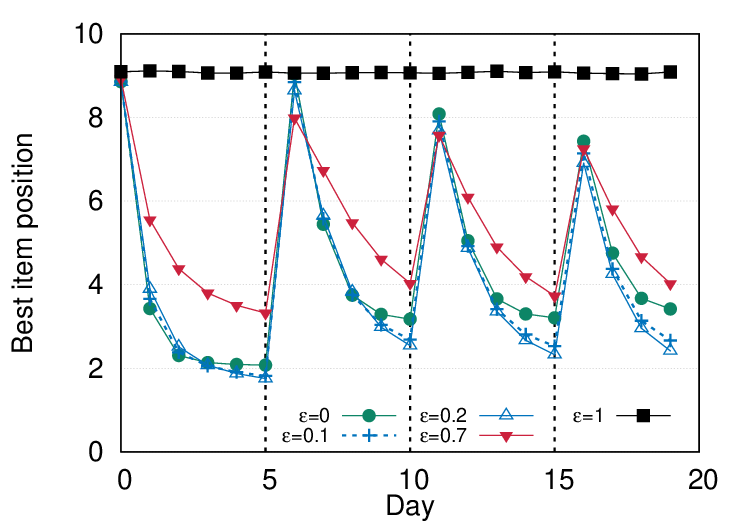}
    \vspace{-0.3cm}
    \caption{Average ranking of the best ride match for a dynamic scenario in which users may change their choice model.}
    \label{fig:dynamic}
\end{figure}

\subsubsection{Dynamic scenario}
\noindent
In this section we consider a \emph{dynamic scenario}, in which each user periodically decides to change his choice model. Specifically, every 5 days a user randomly changes its user category. Figure~\ref{fig:dynamic} shows the variation of the average ranking of the best ride match in the recommendation list. Clearly, after a radical change of the user's choice model, the ranking model provides wrong estimates. However, our learning algorithm quickly detects this change and correctly updates the weights of the ranking function. Interestingly, we can observe that after the first change of user category, the learning algorithm appears slightly slower in updating the user's ranking model. This can be explained by observing that the update rule of the learning algorithm may have some inertia. However, the error introduced by the subsequent changes of user category tends to decrease. As for the static scenario, exploration rates $\epsilon=0.2$ and $\epsilon=0.1$ provide the best learning performance.
%
%
%
%
%
\vspace{-0.1cm}
\section{GoTogether mobile application}\label{sec:prototype}
\begin{figure}[t]
    \centering
    \includegraphics[width=0.4\textwidth]{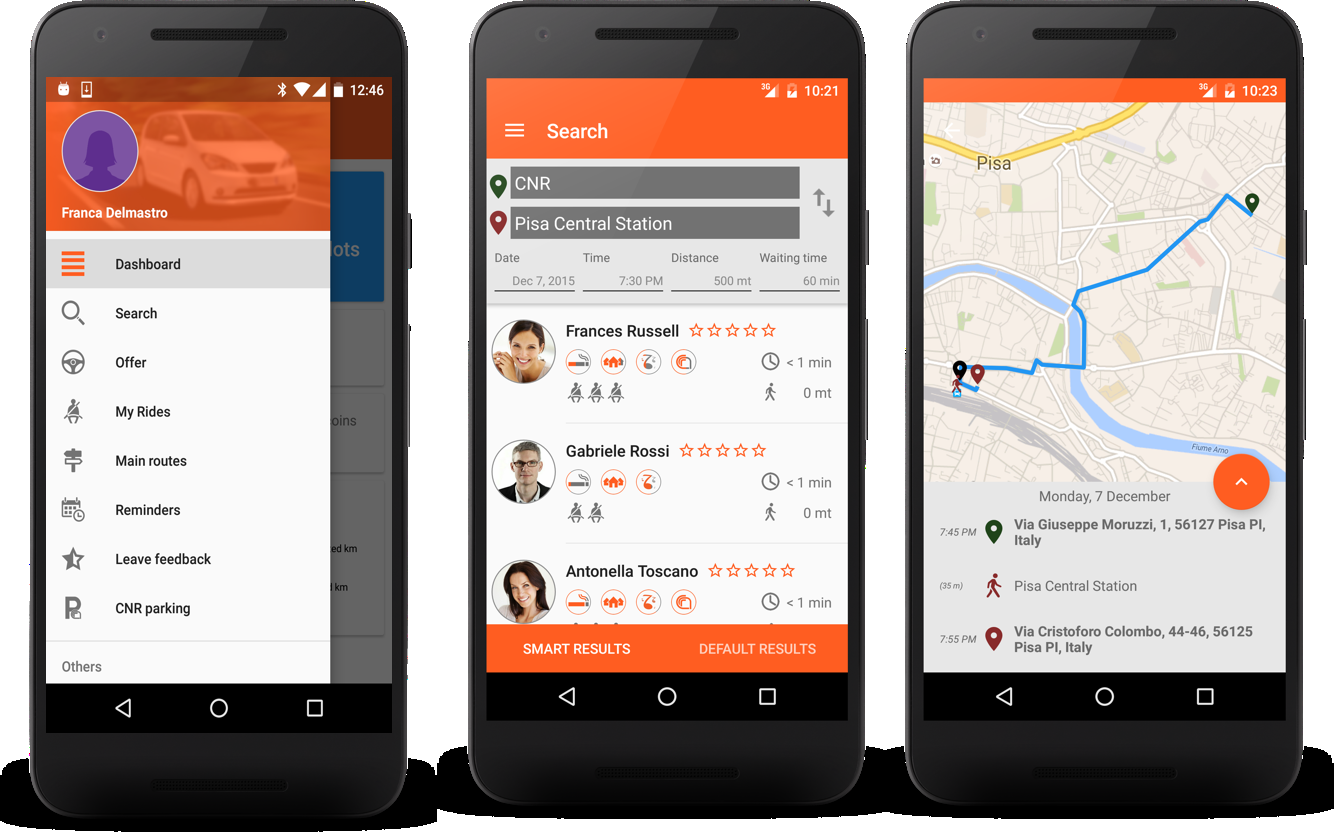}
    \vspace{-0.2cm}
    \caption{GoTogether mobile app.}
    \label{fig:app}
\end{figure}

\vspace{-0.1cm}
\noindent
In order to experimentally evaluate \textsc{GoTogether} with real users, we developed an Android mobile app implementing the recommendation system described above. It has been recently launched in the CNR campus area in Pisa as a corporate carpooling service. The campus hosts more than 1200 working people, several of them commuting every day. The \textsc{GoTogether} app provides several functionalities: ride search and offer operations, visualisation of the current user's rides (both as a driver and a passenger) as well as the most popular shared routes, the possibility to set a reminder to be automatically notified when a plausible trip is available (see Figure \ref{fig:app} for some screenshots). Note that the users' profiles can also be characterised in terms of travel preferences (i.e., listen to music, travel with smokers, colleagues, neighbours), in addition to the temporal and spatial constraints for the requested ride. The application is currently available on the Playstore\footnote{https://play.google.com/store/apps/details?id=it.cnr.iit.smartmobility} and we are collecting real data from its usage to further evaluate the system.

%
%
\section{Conclusions}\label{sec:conclusions}
\vspace{-0.1cm}
\noindent
In this work we have shown that machine-learned ranking techniques can be effectively used to improve the quality of the recommendation system of a car pooling service. In particular, we have designed an online, pairwise learning-to-rank algorithm that leverages on the history of users' selections among the offered rides to predict the individual ranking model of the users. Then, we have used Twitter and Foursquare as data sources to generate a dataset of plausible mobility patterns and ride requests. Finally, we have used this dataset to evaluate our learning algorithm in terms of learning speed and accuracy, both in static and dynamic scenarios. The shown results confirm the validity and robustness of the proposed solution.

As future work, we plan to extend our methodology to consider additional data sources and ride features. Furthermore, we are collecting real data from a prototype implementation of the \textsc{GoTogether} system to evaluate our solution in the real world. Another avenue of research is to design a more sophisticated learning framework that could work in multi-modal scenarios in which car pooling is one of the available on-demand mobility services, in addition to, for instance, car and bike sharing.
\vspace{-0.2cm}

%
%


\bibliographystyle{IEEEtran}

\bibliography{IEEEabrv,carpooling.bib}

\end{document}